# A Proposal for a Fast Infrared Bursts Detector

**Alessandro Drago,**[a,c,1] **Simone Bini,**[c] **Mariangela Cestelli Guidi,**[c] **Augusto Marcelli,**[c,e] **Valerio Bocci,**[d] **Emanuele Pace,**[a,b]

[a] *Dipartimento di Fisica e Astronomia, Università degli Studi di Firenze,*
  *Largo Enrico Fermi 2, 50125 Firenze, Italy*

[b] *Osservatorio Polifunzionale del Chianti,*
  *S.P. 101 di Castellina in Chianti Km 9,25, 50021 Barberino Val d'Elsa (FI), Italy*

[c] *Laboratori Nazionali di Frascati, Istituto Nazionale di Fisica Nucleare,*
  *Via Enrico Fermi 54 - 00044 Frascati (Roma), Italy*

[d] *Sezione di Roma 1, Istituto Nazionale di Fisica Nucleare,*
  *Piazzale Aldo Moro, 2 - 00185 Roma RM, Italia*

[e] *RICMASS, Rome International Center for Materials Science Superstripes,*
  *Via dei Sabelli 119A, 00185 Rome, Italy*

*E-mail*: Alessandro.Drago@unifi.it Alessandro.Drago@lnf.infn.it

ABSTRACT: The gravitational wave GW170817 from a binary neutron star merger and the simultaneous electromagnetic detection of the GRB170817A by Fermi Gamma-Ray Space Telescope, opened a new era in the multi-messenger astronomy. Furthermore, the GRBs (Gamma-Ray Bursts) and the mysterious FRBs (Fast Radio Bursts) have sparked interest in the development of new detectors and telescopes dedicated to the time-domain astronomy across the electromagnetic spectrum. Time-domain astronomy aims to acquire fast astronomical bursts in temporal range between a few seconds down to 1 ns. Fast InfraRed Bursts (FIRB's) have been relatively understudied, often due to the lack of appropriate tools for observation and analysis. In this scientific scenario, the present contribution proposes a new detection system for ground-based reflecting telescopes working in the mid-infrared (mid-IR) range to search for astronomical FIRB's. Experience developed in the diagnostics for lepton circular accelerators can be used to design temporal devices for astronomy. Longitudinal diagnostic instruments acquire bunch-by-bunch particle shifts in the direction of flight, that is equivalent to temporal. Transverse device integrates the beam signal in the horizontal and vertical coordinates, as standard telescopes. The proposed instrument aims to work in temporal mode. Feasibility study tests have been carried out at SINBAD, the infrared beam line of DAFNE, the e+/e- collider of INFN. SINBAD releases pulsed infrared synchrotron light with 2.7 ns separation. The front-end detector system has been evaluated to detect temporal fast infrared signals with 2-12 µm wavelengths and 1 ns rise times. The present contribute aims to be a step toward a feasibility study report.

---

[1] Corresponding author.

**Contents**

**1. Introduction**

A new instrument based on time-resolved infrared (IR) detectors developed for diagnostics in accelerator physics, is proposed for applications in multi-messenger and time-domain astronomy. In the following sections, the proposed device will be introduced alongside the study design for an ultra-fast IR detector suitable for ground-based telescopes. The objective of this article is to present preliminary design concepts and tests to be used for a following, more detailed, feasibility study report. This step could lead to a subsequent Conceptual and Technical Design Report (CDR/TDR). The implementation of the detector in the focal plane of a reflecting telescope will be used to search for astronomical fast infrared bursts (FIRB). Being based on a telescope, the instrument will be also capable of identifying the direction of the source of the bursts, particularly useful in multi-messenger astronomy.

In the section 1, a brief introduction on multi-messenger and time-domain astronomy will introduce the scientific scenario to justify the R&D for the proposed instrument. Basic design concepts are presented in section 2. Test and performance results are described in section 3. Further design considerations for the final implementation are discussed in section 4, and conclusion are reported in section 5.

### 1.1. Multi-messenger astronomy

The main motivations of this work comes from the interest caused by three astrophysical research areas: multi-messenger astronomy, GRBs (Gamma Ray Bursts) and FRBs (Fast Radio Bursts), the latter being both fast transient phenomena.
1n 1987, the supernova SN1987A event with electromagnetic and neutrino emission was the first multi-messenger observation. The gravitational wave GW170817 [1] caused by a binary neutron star merger and the simultaneous electromagnetic detection of the GRB170817A by Fermi Gamma-Ray Space Telescope, opened a new era in multi-messenger astronomy [2]. Other electromagnetic observations correlated to the GW170817 event, were detected in parallel by several telescopes in different electromagnetic ranges, too. The signal lasted for approximately 100 seconds starting from a frequency of 24 Hz, increasing in amplitude and frequency to a few hundred Hz.
It is interesting to note that the previous gravitational wave event GW150914 [3] caused by the coalescence of two black holes did not display clear multi-messenger detection. In this event, the chirp signal lasted over 0.2 seconds, and increased in frequency and amplitude in about 8 cycles from 35 Hz to 250 Hz.
First of all, in terms of the instrument design, it should be highlighted that the nature of gravitational wave detection is given by one-dimensional (1-D) signals versus time. In the field of accelerator physics, temporal diagnostic devices are called "longitudinal" [4]. Gravitational wave observatories like LIGO and VIRGO are devices that intrinsically work temporally. In the signal theory, this is equivalent to recording 1-D signals. Hence, the gravitational wave observatories in operation correspond to three 1-D signals in total, two from the LIGO locations in USA, and one from VIRGO in Italy. Two other GW observatories are planned in India and Japan and can help to make more directionally precise the detection. Analysis of the slight variation in arrival time of the GW at the detector locations yielded an approximate angular direction to the source, but a multi-messenger detection strategy can be more accurate.

### 1.2. GRB and FRB

Gamma Ray Bursts (GRBs) and Fast Radio Bursts (FRBs) are other fast transient phenomena observed. The former one can be detected by the Fermi Gamma-Ray Space Telescope (FGRST), a satellite orbiting in the Low Earth Orbit (LEO) built to detect the gamma-ray astronomy. It is made up of two instruments: the Large Area Telescope (LAT) [5], a gamma telescope mainly designed for imaging, and the Gamma-ray Burst Monitor (GBM) [6] designed to detect fast gamma-ray bursts. The Fermi-GBM has detected the gamma-ray burst GRB170817A temporally related to the GW170817. GRBs were either labeled short, long or very long. Short GRBs can last milliseconds, usually around 0.2 s, while long GRBs can last more than 2 s [7]. There is a general consensus about the origin of GRB events: these brief flashes of high-energy light result from some of the universe's most explosive events, including births of black holes, supernovae explosions and collisions between neutron stars [8].

A FRB, on the other hand, is a radio frequency burst with a duration ranging from a fraction of a millisecond to 3 seconds. Many oscillate around 1400 MHz; some have been detected at lower frequencies in the range of 400–800 MHz [9] [10]. No dominant theory has emerged about the sources of FRB in recent literature, where many different interpretations of these phenomena have been proposed [15].

### 1.3. Time-domain astronomy

Recent literature has highlighted the utility to developing instruments for the time-domain astronomy with the goal to acquire fast astronomical bursts or transients with duration from a few seconds to 1 ns in all electromagnetic ranges. Among the others, the interest was underlined by Judith Racusin (NASA) in a recent talk about the Fermi Space Telescope mission [11].

In this research area, another experiment with observations in time-domain from many astronomical objects, has been carried out at the LICK Telescope, in California, in the NIR (Near InfraRed) range by a team from the Center for Astrophysics and Space Sciences, University of California, San Diego [12].

Of course astronomy has always been an observational science working in the time-domain, but, in classic visual astronomy, the time scales range from minutes to millennia. On the other hand, modern ground-based or satellite telescopes work by integrating signals. They acquire images, often extremely detailed ones, such as those produced by the Hubble Space Telescope (HST) [13] or the James Webb Space Telescope (JWST) [14]. MIRI (Mid-IR Instrument) is the only module of the JWST that can operate at mid-infrared wavelengths, offering the following modes: imaging, coronagraph and spectroscopy over 5 µm−28 µm wavelength range. For photometric measurements, detectors need time to integrate signals.

Considering that time-domain astronomy spans over time scales from a few seconds to 1 ns (2-3 s to $10^{-9}$ s), dedicated very fast detectors for telescopes need to be designed. Furthermore, different types of systems are necessary to observe the electromagnetic spectrum in different ranges. This project is focused on the mid-infrared (mid-IR) part of the electromagnetic spectrum, from 2 to 12 µm of wavelength.

This range was chosen ruling out gamma-ray frequencies, given that they already benefit from several ongoing experiments and well-understood astronomical principles. Furthermore, we ruled out the radio wave spectrum. FRBs detected within these frequencies are linked to phenomena that remain subjects of ongoing debate [15] and as such radio telescopes are numerous and well-funded, like for example FAST (Five-hundred-meter Aperture Spherical radio Telescope) in China, the most recent observatory. In the mid-IR range, on the contrary, there are few experiments for time-domain detection in the current research and it seem interesting to investigate if in this frequency range it is possible to observe astronomical fast bursts.

### 1.3 Directionality of detection

The Einstein Telescope [16], a proposed third generation GW observatory, is explicitly designed with features to identify better the coordinates of the GW source. The importance of this is so great to condition deeply the very design of the observatory, based on interferometers with two L-shaped arms, or on a triangular scheme with three arms.

Considering the interest to identify the direction of the bursts, it would be useful to have a direction pointing device.

Summarizing, the basic points to justify a new type of telescope, are:
- Why look for fast IR bursts ? Answer: after gamma and radio, it is interesting to observe fast short signals in other parts of the electromagnetic spectrum.
- Could fast IR bursts exist ? Answer: specialized tools are necessary to answer the question.
- Are there theories that predict them? The numerous theories proposed to explain Fast Radio Burst sources can also be largely adapted to the Fast IR Bursts [15].
- How to detect them? The aim of this article is to propose a novel instrument to search for fast astronomical bursts, transients or flares in the mid-infrared range.
- The proposed instrument is inherently directional, which is important for identifying burst sources.

## 2. Design outlines

### 2.1. Enrico Fermi's accelerator laws

Enrico Fermi was the first scientist to describe the ISM (Inter Stellar Medium) as an accelerator of charged particles. He proposed mechanisms that now are the two Fermi's accelerator laws.

The *First Order Fermi Acceleration Law* is based on "shock waves". If a particle encounters a moving change in the magnetic field, this can reflect it back through the shock (downstream to upstream) at increased velocity. If a similar process occurs upstream, the particle will gain energy again. These multiple reflections greatly increase its energy.

The *Second Order Fermi Acceleration Law* describes the gain in energy obtained by a charged particle while moving in the presence of "magnetic mirrors" that are moving randomly. If the magnetic mirror is moving towards the particle, the particle will end up with increased energy upon reflection. The opposite holds if the mirror is receding. This notion was used by Fermi in 1949 to explain the formation of cosmic rays.

Hence the idea of applying the technology of diagnostic devices used in circular accelerators to a new instrument for astronomical observations for the detection of fast bursts and transients.

### 2.2. The synchrotron radiation infrared beamline

In order to design a time-domain astronomical detector in the mid-infrared range, the experience acquired in the lepton beam diagnostics for circular accelerators has been applied. The new detection system has been tested at SINBAD [17], the infrared light beamline [18] at DAFNE [19], the e+/e- circular collider of the National Laboratory of Frascati of INFN. The accelerator is composed by two main rings having 97 m length, 510 Mev energy, 368 MHz radio frequency and 120 electromagnetic buckets. DAFNE has two interaction regions where positron and electron bunches can collide at an energy of 1.02 GeV in the center of mass. The main rings have some synchrotron light beamlines with different energy available for experiments. SINBAD is the IR beamline where pulsed infrared light with ~ 0.5 eV energy is emitted by the accelerated electron beam. The photons exit from a bending electromagnet through an exit port connected to a pipe of few meters. The electron beam is split in bunches by the radiofrequency system and, as consequence, the synchrotron light is pulsed every 2.7 ns. Usually, there is also at least a gap in the e- bunch train to avoid the ion trapping effect. See in Fig. 1 the final part of the beamline with the infrared light spot on the mirror on the left.

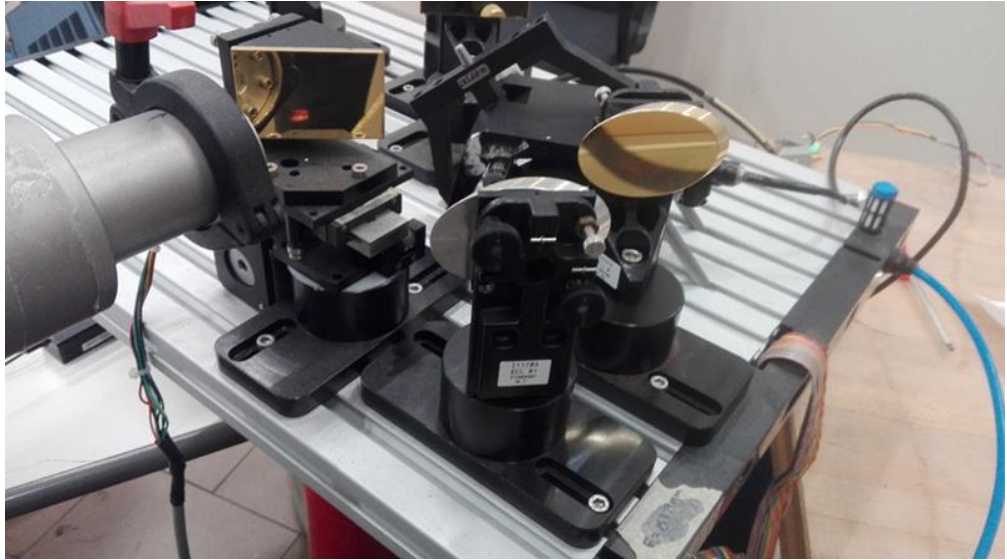

Fig. 1 Final optical layout with three mirrors at SINBAD beamline. The IR spot is visible on the mirror on the left.

*2.3. Diagnostics for circular accelerators*

To be in operation, a circular accelerator needs many diagnostic devices. Some devices work in transverse and other in longitudinal. In the jargon of accelerator physics, beam diagnostic devices integrating over time to take vertical and horizontal images of the beams are called transverse, while those acquiring in time-domain are called longitudinal (temporal). For the former devices, the phase difference or the time delay of the charged particle bunches are referred to the master clock of the accelerator that has picosecond precision.

It is possible to see the images in the visible of the electron or positron beams from the synchrotron light monitor (SLM), that works transversally, in Fig. 2. The SLM does not show any evidence of bunches.

In Fig. 3, a high frequency oscilloscope shows the two stored beams versus time, the positrons in red and the electrons in blue. Two electromagnetic pickups are connected to the instrument and it is possible to see the bunch separation made by the radio frequency system. Indeed to restore the energy lost in every turn by the charged beam, a high voltage kick by the radio frequency system is applied to a copper cavity. The radio frequency working at 368 MHz breaks the e- beam in buckets with 2.7 ns distance. In Fig. 4 a zoom of the same plot is shown.

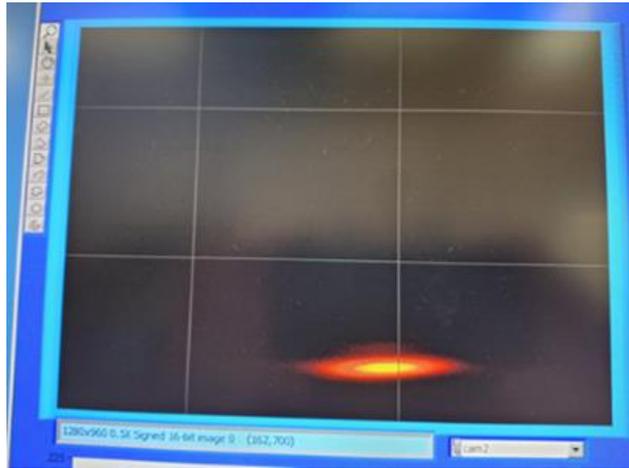

Fig. 2 DAFNE e- synchrotron light monitor (SLM) in the DAFNE control room. The SLM is a transverse device integrating the visible radiation by the electrons. It does not show any evidence of bunches.

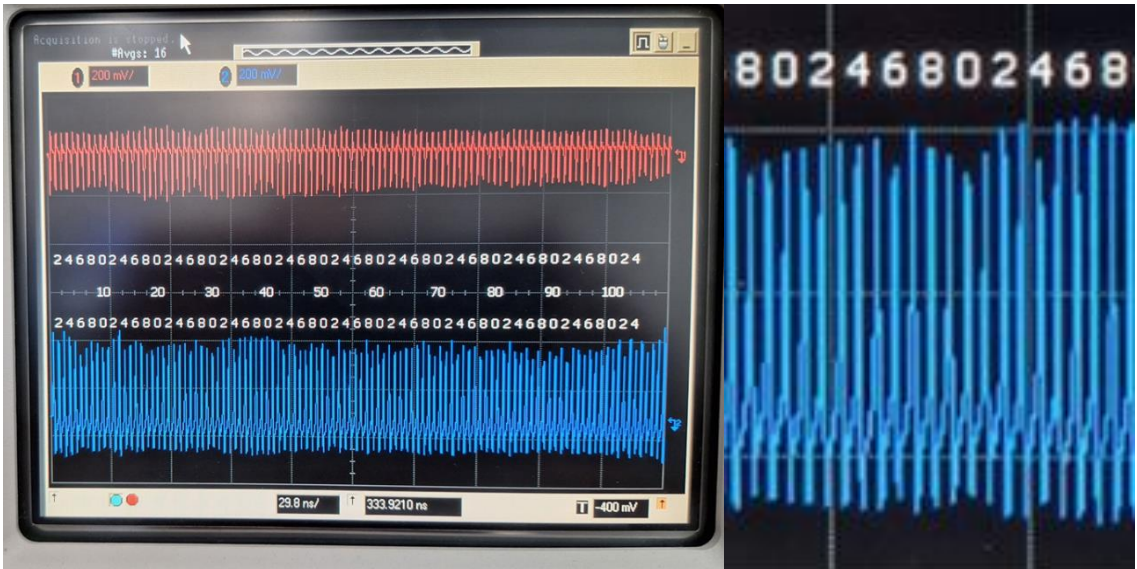

Fig. 3 In the left, positron (red) and electron (blue) 110 bunches longitudinal detection (in time-domain) by two electromagnetic pickups connected to an oscilloscope, showing the separation of the bunches. Fig. 4 In the right, a picture zoom of the e- bunches. The plot shows the amplitudes, in vertical, versus the time, in horizontal.

*2.4. Design activities*

Time-resolved detectors in the mid-IR have developed at LNF for the beam diagnostics for the e-/e+ storage rings of DAFNE with positive results [20] [21]. From the previous experience, a detector has been designed for ground-based directional observations by using a reflecting telescope. The implementation for scientific ballons has been considered, too. Looking to these goals, the best choice is for semiconductors working at room (or lower) temperatures with rise time and fall time of the order of the nanosecond. The detection system should be as compact as possible to be used as mobile or transportable device to be easily put in the telescope focal plane. The HgCdTe detectors (also called MCT, Mercury Cadmium Tellurium) are photoconductors made by the VIGO Photonics S.A. [22]. They have been chosen given that they accomplish the required specifications [23] in term of detectivity and frequency response at room temperature. The choice of photoconductors gives faster rise time and fall time than by implementing photovoltaic devices. However, using photoconductors, a transimpedance circuit is necessary to adapt the electrical signal from current to voltage. The conversion is necessary for using voltage amplifiers. A custom electrical circuit has been designed and fabricated to have a small 4-layers PCB (Printed Circuit Board), easy to be interfaced with a classic reflecting telescope as Ritchey-Chrétien or Cassegrain, by installing the detector board in the focal plane. The board, with the dimensions of a cellular phone (7x14 cm), is shown in Fig. 5 and Fig. 6. The PCB can host up to 19 pixels with TO-39 case, a transistor common case.

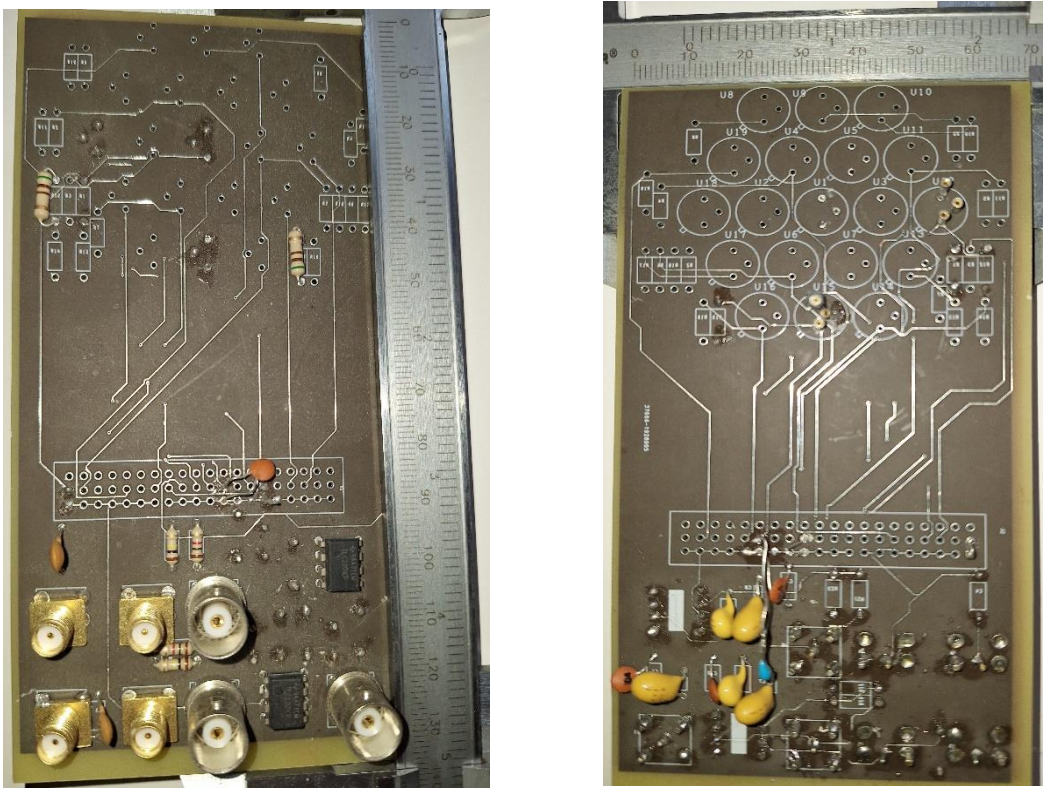

Fig. 5 and Fig. 6: Four layers printed circuit, front and back sides (14x7 cm).

Four HgCdTe (MCT) photoconductors and one InAsSb photovoltaic semiconductor have been used during the tests. A high frequency bias-tee has been used for the transimpedance circuit necessary to interface the voltage amplifiers. The detectors tested are the following:
- PC10.6 with TO-39 case by Vigo, single pixel, 50x50µm active area.
- PC10.6 R005 with BNC case by Vigo, single pixel, 50x50µm active area.
- PC10.6 R020 with BNC case by Vigo, single pixel with 200x200µm active area.

- InAsSb photovoltaic device by Hamamatsu.

The latter is an alternative to the MCT photoconductors, a cheaper InAsSb detector. Being photovoltaic, it does not need the transimpedance circuit that work as high pass filter. The photovoltaic can be easily used at low frequency, even at extremely low frequencies (µHz).

A custom printed circuit board designed at LNF, can allow the placement of up to 19 pixels, all working in time-domain mode, that can be all connected in parallel. Each pixel corresponds to a Vigo PC-10.6 MCT photoconductor, with TO-39 package, while the BNC type package is non compatible. The Hamamatsu detector uses a similar transistor-type package, it can be loaded on the same custom PCB and does not need current. Seven pixels, in the center of the board, are used to acquire the real astronomical signal and other twelve semiconductors, to be covered, are used to record the dark signal that has to be subtracted in real time during the acquisition. The module outputs are six: DC/low frequency, high frequency and high impedance for both the real signal and the dark signal. The other components are the bias-tee model ZFBT-4R2G by Mini-Circuit, having 0.1-4200 MHz bandwidth, and three RF amplifiers used in alternative: the ZFL-500 and the ZFL-1000LN by Mini-Circuit and a cascode amplifier designed at LNF. In addition, an operational amplifier, the Texas Instruments TLC2201, was used for low frequency range acquisitions [24][25]. The most relevant results of the tests are discussed in the section 3.

### 2.5. Design blocks

To implement the final acquisition system it is necessary to design a series of blocks that have to be engineered. The first three elements in the following list, will be part of the front-end board, completely analog, while other elements are related to the digital parts. In the following there is a list of main blocks and items. Note that the blocks a) b) c) are part of the current tested hardware. The block d) has been tested only partially because, as data acquisition system, an oscilloscope has been used with not variable sampling frequency and poor storage capability. About the point j), the off-line emulator and classifier software has been developed and it is working. This is the list of the blocks:

a) Detectors: two signal inputs are foreseen, one from the real IR light transducer(s) and the other from the dark light transducer(s) having a cover on the components. The detectors are the HgCdTe tested in laboratory. However also the other technology, the InAsSb photovoltaic detector by Hamamatsu Photonics, has been evaluated for managing low frequency transients.

b) Transimpedance circuits are necessary if the MCT components are photoconductors. Photovoltaic detectors do not need the transimpedance circuit. The transimpedance circuit cuts the low frequency part of the signal.

c) Amplifiers: both low and high frequency devices have been tested. Some of them are designed in the laboratory and others are bought from commercial distributors. Low noise devices are necessary. An amplifier covering all the bandwidth of the detector has not been found.

d) Analog-to-digital converters (ADC): almost two, but preferably four, will be needed. Two ADC's for the signals, one for the trigger, one for a test input. ADC with 16 bits are preferable.

e) ADC sampling clock: the clock frequency will be chosen by the operator. A slower clock can be easier to manage for long acquisition traces of signals with low frequency bandwidth. A faster clock, for technological and practical reasons, limits the temporal length of the records but it is necessary if the frequency bandwidth of the signal is higher. To acquire signals with 1 ns rise time, 4 GHz analog bandwidth and 10 Gsamples/s are necessary to overcome the sampling theorem of Shannon-Nyquist.
   f) Acquisition and processing of both signals: real astronomical light and dark light. This software module can be embedded inside an oscilloscope or implemented in a dedicated data acquisition system.
   g) Registration to memory storage of both analog inputs, real signal and dark signal, and time stamp, equatorial coordinates of the pointing system, ADC clock frequency.
   h) Display for visualization in time-domain and in frequency domain.
   i) Acquisition triggers as described in the following par. 4.1.
   j) Off-line FIRB emulator and classifier, see par. 4.3.
   k) Other items: Power supply: 12Vdc and/or 220Vac and compact portable containers for the different setups are necessary.

The detection system will need to be tailored for different setups, whether it is used for ground-based telescopes or for acquisition by using stratospheric balloons. There are differences between these two situations and the detection system designs will vary significantly.

The low and the high frequency amplifiers have different characteristics that impact on the ADC's and the low-pass filters. As a consequence, the digital acquisition may have different sampling clocks, going from slow (few tens of kHz) to fast (10 GHz) respectively. The low-frequency system can be implemented with a cheaper technology, smaller memory and lower electrical power requirements.

The setup for a stratospheric balloon must be implemented in a very compact box including the front-end analog board as well as the digital parts with the ADC's, processing module and memory storage. A heater must be included to maintain the box temperature > 0 degrees. A communication system is not foreseen and, for this reason, the acquisition clock frequency and the threshold trigger values have to be selected before the launch. Telescope and visualization systems are not included in the setup for balloon flights.

## 3. Tests and performance
### 3.1. Tests carried out at LNF

In May 2022, preliminary tests done by using two photoconductors of the final type, even if with BNC type package, have been tested at SINBAD with very promising results: the PC-10.6 R005 model and the PC-10.6 R02. Also another setup has been used to test the photodetectors by using pulsed IR light emitted by a LED driven by a signal generator. A commercial photodiode has been used, the IR LED model OSLON Black, emitting at 940 nm, with 90 degrees of aperture and 320mW/sr of power. To produce pulsed IR light, the photodiode is powered directly by a pulse generator. The light made by the LED is just on the low limit of the photoconductor frequency band, but it works. However the setup has slower rise time and fall time respect to the synchrotron light pulses. After the preliminary tests in the laboratory, the hardware has been moved to the SINBAD beamline for activities before the DAFNE shutdown of mid-July 2022.

For the tests at the SINBAD beamline [26], a 4-channel oscilloscope was used to acquire the signal from the train of 110 bunch of electrons. The oscilloscope inputs were connected as in the following:
1) a signal from an electromagnetic pickup in the e- main storage ring;
2) the pulsed IR synchrotron light signal transduced in electric and amplified;
3) a "dark light" signal transduced in electric and amplified;
4) the DAFNE revolution master clock used as trigger for all the input signals.

Considering that the signals come from a storage ring, they are repetitive every 324 ns and their difference in phase, due to different paths, is not meaningful. The signal 1) from the electromagnetic pickup is necessary for a real time comparison with the particle beam to check the correct alignment of the detector placed in front of the SINBAD output port. The detector must be correctly put on the focal plane to produce the electric signal 2). The "dark light" signal 3) is used to evaluate and subtract the environmental and instrumental noise present in the room where the detection system is located. However in the situation described, the dark signal acquired has shown extremely low amplitude. It is important to note that the real IR light signal 2) can be affected by other sources of noise: a) the environmental and instrumental noise shaking the e- beam in DAFNE hall and producing similar effects on the emitted synchrotron photons and b) the thermal noise inside the HgCdTe semiconductor. Both latter noises will not have impact on final setup for astronomical observations because the noise coming from the e- beam will not be present and the thermal noise will be mitigated by the low temperatures typical of the nocturnal observations.

Even if the devices are declared by the VIGO company with a responsivity peak at 10.6 µm, the real sensitivity peak seems between 7 and 9 µm from the producer data sheet. However the shape of the response curve is spread from 2 to 12 µm.

The detector R02, with the active area larger than the R005, has produced an 3 mV peak-peak output, while the R005has produced a 2 mV peak-peak output at the same bunch current. The explanation is simple: the IR light spot in the focal plane is slightly larger than 50 x 50µm active area of the R005.

During all the year 2022 tests, a loud noise was present on the IR signal, that, if real, would be not compatible with a detector for telescope. Managing repetitive signals as those from a circular accelerator ring, the noise has been cancelled by integrating the signals and averaging the acquisition of 10k samples. In the 2022, the source of the problem was not clear, however, in the tests done in the electronic laboratory, this noise was not present.

In April 2023 after the start of a new run of DAFNE, the source of the low frequency noise was identified. The shake of the beam was real. It came from the DAFNE radio frequency (RF) system which was faulty and eventually completely repaired. More in detail the mode-0 RF feedback working badly provokes a shaking in energy of each bunch with a frequency in the interval 1-30kHz, where the synchrotron frequency is 30kHz.

This positive conclusion demonstrates the correct real time behavior of the HgCdTe detector and the possibility of using it for astronomical observation. In conclusion, the detector has correctly acquired the shaking of the buckets and the pulsed synchrotron light and does not need to averaging/integrating the signal.

*3.2. Detector performance*

To evaluate the performance versus temperature, tests have been carried on to compare the behavior of the VIGO detectors first at room temperature (RT) and then at lower temperature. This was done using two Peltier cells as a sandwich with the device in the middle. No difference was found in the noise level, showing that cooling is not necessary. As said above, using the pulsed light from SINBAD, four different detectors and five different amplifiers have been evaluated. The HgCdTe photoconductors, with different active area or package, have shown excellent performance with fast light pulses with 1 ns rise time. A drawback is that the photoconductor devices need a transimpedance circuit that rules out the use of the detector at low frequency, under ~10k Hz.

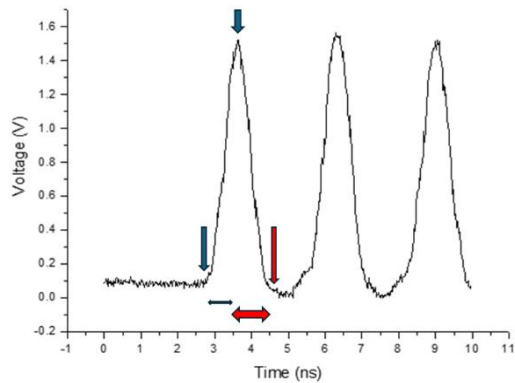

Fig. 7: Transduced IR light signal of three e- bunches measured by the IR photoconductor R005. Rise time and fall time are about 0.7 and 1 ns, respectively. Bunch current was ~14 mA.

For comparison, a photovoltaic detector, based on a different and cheaper technology, a InAsSb semiconductor, has been tested. Being photovoltaic, it does not require the transimpedance circuit and, for this reason, it can work at extremely low frequency. As drawback, the photovoltaic technology shows less sensitivity, about one third of the photoconductor, and more noise (10-20% more). The smaller signal response has requested the use of two 35 dB amplifiers for a total of 70 dB voltage gain.
The evaluation of the detector sensitivity has been done by using the IR light of the SINBAD beamline. The photon source comes from a bending magnet with a radius of 1.4m and a magnetic field of 1.2T. The energy of the emitted photons has a large spread, well outside the detector sensitivity that is declared from 2 to 12 µm.
The photon beam divergence is 17 mrad (horizontal) x 35 mrad (vertical). The photon flux (calculated) is $10^{13}$ photons/s x 0.1% bandwidth [27] [28]. The SINBAD photon flux is calculated for the total beam current of 2 A. However, the detector is sensible to the IR pulses emitted by each electron bunch, not to the total beam current. Some calculation is necessary. The DAFNE harmonic number is 120, but to avoid the ion trapping effect, some buckets are left empty, storing usually in the ring a train of 100 bunches followed by a gap. Hence each bunch emits $10^{13}/100 = 10^{11}$ photons per second. The $10^{11}$ ph/s value must be divided by the revolution frequency that is 368 MHz /120= 3.067 MHz because the measurement is done for one bunch in one turn. The result is 32609 photons emitted per one bunch of 20 mA in each turn:

photons@20mA     = 32609 photons   (photons = flux/n_bunch/freq_revolution )
photons@10mA     = 16304 photons   (= photons@20mA/2)
photons@2mA      = 3261  photons   (= photons@20mA/10)

The measurements are done during the e+/e- collisions with bunch current between 2 and 10 mA. Given that, in this range of bunch current, the detector signal is correctly displayed and measured (peak-peak voltage) by the oscilloscope, the performance in term of sensibility to the pulse is tested between 3.2k and 32k photons x 0.1% bw ($\delta\lambda/\lambda$) without saturation or distortion.

For a bunch of 2mA, a signal of 2 mV peak-to-peak has been measured after 20 dB voltage amplification, that corresponds to 0.2 mVpp detector output without any amplification.

Considering the realistic possibility, for the engineered system, to use 80 dB amplification stage, that it would be equivalent to a 10k linear gain, the detection system shall be sensible to 1 photon for 0.1% of the infrared bandwidth ($\delta\lambda/\lambda$) by using the needed amplification.

In Fig.7 the three bunches of the electron beam are measured by the uncooled IR photoconductor detector R005. The rise time and the fall time of the IR signals are about 0.7 and 1 ns, respectively. Bunch current was ~14 mA. The signal is acquired from the pulsed IR light at the SINBAD beamline, converted in electric by the HgCdTe detector and amplified. To make the measurement in Fig.7, the signal has been amplified by 40 dB voltage amplifiers with bandwidth 10kHz-4GHz.

About the InAsSb detector, similar results have been obtained with 20% more noise and increasing the voltage amplification to 70 dB.

### 4. Further design considerations for an actual implementation

*4.1. Triggers for the data acquisition*

Triggers are necessary for recording only interesting data to be analyzed off-line. Summarizing, three types of triggers to the data acquisition can be used:
   a) The accelerator master clock from the DAFNE timing system during the data taking at the SINBAD beamline.
   b) The sync output from a pulse generator for self-diagnostics.
   c) A smart trigger generated by a threshold module making a comparison between the real astronomical signal and the dark signal. The threshold level should be extrapolated in base at environmental noise and temperature noise levels acquired by the dark signal. The trigger threshold value, computed from the dark signal, must be averaged and updated every 30 seconds. This is to avoid to store not meaningful data. A variable gain must be foreseen to be applied by the human operator for increasing the system flexibility.

*4.2. Critical aspects*

There are, however, critical aspects that have to be considered because they can limit the observations. For ground-based telescope implementation, the list of critical issues is as follows:

   a) The temperature must be in the range $0 < t < 35$ Celsius degrees for the data acquisition system to work. A larger range of temperature is possible but it will impact on the cost requiring special components.
   b) The seeing should be not a big issue. It can cause some jitter or distortion of the signal.
   c) The atmosphere offers good infrared transmission windows only between 8 and 14 μm and between 3 to 5 μm. This can cause some loss to the signal, jitter or distortion.
   d) The weakness of the astronomical signals can be under the sensitivity of detector. Implementing a PCB with 7 pixels in parallel can help. Increasing the amplification stage can be also considered.

*4.3. Burst classifications by A.I. techniques*

The data acquisition system is completed by implementing artificial intelligence techniques. This is necessary because the astronomical IR transients could be most likely extremely rare (few events per year). The trigger will select events that have to be recorded together with observation data and environmental conditions. The transient tracks will be stored to be analyzed off-line by the inference engine and classified by measuring features and expressing probability values. The FIRB emulator and the classifier are implemented by procedural approach software (Matlab/Octave). In the future the classifier can be implemented by artificial neural networks (ANN). The test of the FIRB classifier need a data base made by the emulator. This program has created a data base containing patterns with different value of grow rates, shapes, noise and jitter. After the creation, the database is analyzed by the classifier. The procedural approach is based on "syntactic" rules to classify the shape of the signals, while the ANN will require a Machine Learning (ML) training phase before starting the analysis of the real tracks. The ML training will be based on the database generated by the emulator.

**5. Conclusions**

A new ultra-fast mid-infrared detector for ground-based telescopes has been proposed with the objective to search for astronomical FIRBs (Fast mid-InfraRed Bursts). It is a directional instrument, and this feature can be useful for observations in the areas of multi-messenger and time-domain astronomy.
The main elements of the instrument are:
1) a reflecting ground-based telescope (Cassegrain or Ritchey–Chrétien) dedicated to the experiment for long periods of time;

2) an infrared detector for fast signals with rise time of the order of one nanosecond to be put in the focal plane of the telescope; the detector converts mid-IR light to electric analog signal;

3) analog circuits (bias-tee, pcb to put in parallel 7+12 detectors, amplifiers) to adapt the analog signal for the digital acquisition system;

4) a digital acquisition system sampling at 4GHz and storing the selected data;

5) a smart trigger to record only valid and interesting events;

6) off-line programs with A.I. techniques to analyze the records without human operators. Points 2) and 3) are the analog front-end of the experiment and are the most critical points and they have been tested. Points 1) and 4) require a big budget (also for the use of a powerful telescope), not R&D activities. Points 5) is under development. The point 6), off-line software, to emulate and to classifier bursts, has been developed.

A preliminary design for a fast infrared telescope with an evaluation of the performance has been discussed. The analog front end has been tested including detectors, bias-tee, a custom printed circuit board for allocate up to 19 pixels, and different amplifiers. From tests and calculations, the analog front end, based on HgCdTe photoconductors, if equipped by 80 dB voltage amplification with 4 GHz bandwidth, shall be sensible to 1 photon for 0.1% of the infrared bandwidth ($\delta\lambda/\lambda$).

It is a step toward a feasibility study. Astronomical tests should be arranged to go in detail in the design. The goal is to place the mid-IR detector in the focal plane of a ground-based reflecting telescope, as a Cassegrain or a Ritchey-Chretien. The 80 cm Ritchey-Chrétien telescope of the Osservatorio Polifunzionale del Chianti [29] is a possible option. This article describes the preliminary stage for funding request necessary to make progress toward a complete CDR/TDR.

**Acknowledgements**
Thanks to all the SINBAD team for supporting the tests performed at the synchrotron radiation beamline at DAFNE-L. Thanks to the LNF vacuum laboratory for support in electronics and instrumentations. The FAIRTEL (Fast InfraRed TELescope) experiment was funded by the CSN5 (Commissione Scientifica Nazionale 5) of the INFN.

**References**
[1] B. P. Abbott et al. (LIGO Sc. Coll. and Virgo Coll.), *GW170817: Observation of Gravitational Waves from a Binary Neutron Star Inspiral*, *Phys. Rev. Lett.* 119 (16): 161101.95. (2017). https://doi.org/10.1103/PhysRevLett.119.161101. https://arXiv.org/abs/1710.05832.
[2] B.P. Abbott et al. (LIGO, Virgo and other collaborations), *Multi-messenger Observations of a Binary Neutron Star Merger*, *Astroph. J.* 848 (2): L12. (2017). arXiv:1710.05833. DOI:10.3847/2041-8213/aa91c9 **.**
[3] B. P. Abbott et al. (LIGO Scientific Collaboration and Virgo Collaboration), *Observing gravitational-wave transient GW150914 with minimal assumptions*. *Phys. Rev. D* 93, 122004, Pub.7 June 2016; *Erratum Phys. Rev. D* 94, 069903 (2016). https://doi.org/10.1103/PhysRevD.93.122004
[4] Shyam Prabhakar 2001, "New Diagnostics and Cures for Coupled-Bunch Instabilities". PhD Thesis, Stanford Un.. SLAC-R-554, SLAC Report.

[5] W. B. Atwood et al., *The Large Area Telescope on the Fermi Gamma-Ray Space Telescope Mission*, Astrophys. J. 697, pp. 1071-1102. https://doi.org/10.1088/0004-637X/697/2/1071

[6] E. Bissaldi et al., *High-redshift Gamma-Ray Burst Studies with GLAST*, In: *Gamma-Ray Bursts: Prospects for GLAST* (M. Axelsson and F. Ryde, eds.), vol. 906 of Am. Inst. of Ph, Conf. Proc., pp. 79–88, May 2007. https://doi.org/10.1063/1.2737409

[7] E. Le Floc'h et al., *The first Infrared study of the close environment of a long Gamma-Ray Burst*, Astrophys. J. 746 (1): 7. (2011) https://arXiv.org/abs/1111.1234 https://doi.org/10.1088/0004-637X/746/1/7

[8] B. Abbott, et al. 2008, "Implications for the Origin of GRB 070201 from LIGO Observations". 2008. The Astrophysical Journal, Volume 681, Number 2
Citation B. Abbott et al 2008 ApJ 681 1419. DOI https://doi.org/10.1086/587954

[9] D. Lorimer et al., *A bright millisecond radio burst of extragalactic origin*. Australia Telescope National Facility. *Science* 2 Nov 2007, Vol 318, Issue 5851, pp. 777-780, DOI: https://doi.org/10.1126/science.1147532
https://www.science.org/doi/abs/10.1126/science.1147532

[10] The CHIME/FRB Coll., *A second source of repeating fast radio bursts*, Nature 566 (7743): 235–238. (2019). arXiv:1901.04525. Bibcode:2019Natur.566..235C. https://doi.org/10.1038/s41586-018-0864-x  PMID 30653190. S2CID 186244363.

[11] J. Racusin (NASA), *Surveying the Dynamic Universe with the Fermi Gamma-ray Space Telescope*, L'Aquila Joint Astroparticle Colloquium, 4/27/2022.

[12] J. Maire et al, *Search for Nanosecond Near-infrared Transients around 1280 Celestial Objects*, *The Astronomical Journal*, Vol.158, Num.5. 2019 AJ 158 203. https://iopscience.iop.org/article/10.3847/1538-3881/ab44d3

[13] A. G. Riess et al., *A Redetermination of the Hubble Constant with the Hubble Space Telescope from a Differential Distance Ladder*. Astrophys. J., Volume 699, Number 1. 2009. https://doi.org/10.1088/0004-637X/699/1/539

[14] R. Schoedel et al., *The JWST Galactic Center Survey. A White Paper*, https://arxiv.org/abs/2310.11912  [astro-ph.GA] 2 Nov 2023.

[15] https://frbtheorycat.org/

[16] M. Branchesi et al., *Science with the Einstein Telescope: a comparison of different designs*. Journal of Cosmology and Astroparticle Physics, Vol.2023, July 2023. DOI 10.1088/1475-7516/2023/07/068

[17] A. Marcelli et al., *Infrared beamline SINBAD at DAFNE: expected performance at the sample site*, *Proc. of SPIE*, Vol. 3775, 7. *Accelerator-based Sources of Infrared and Spectroscopic Applications*; (1999) https://doi.org/10.1117/12.366639

[18] http://dafne-light.lnf.infn.it/

[19] C. Milardi et al., *DAΦNE RUN FOR THE SIDDHARTA-2 EXPERIMENT*, presented at the *IPAC'23, 14th International Particle Accelerator Conference*, 7-12 May 2023, Venice, Italy. https://accelconf.web.cern.ch/ipac2023/pdf/MOPL085.pdf

[20] A. Bocci et al., *Fast Infrared Detectors for Beam Diagnostics with Synchrotron Radiation*. Nucl.Instrum.Meth.A 580:190-193. 2007. DOI: 10.1016/j.nima.2007.05.081

[21] A. Drago et al., *Fast Rise Time IR Detectors for Lepton Colliders*". *4th ICFDT 2016*, *JINST* 11 no.07, C07004, https://doi.org/10.1088/1748-0221/11/07/C07004

[22] https://vigo.com.pl/en/home/ ,  https://vigophotonics.com/

[23] J. Piotrowski, M. Galus and M. Grudzien, *Near room-temperature IR photo-detectors*. Infrared Phys., 31(1), 1−48. (1991).

[24] A. Drago et al., *Ultra-Fast InfraRed Detector for Astronomy*. Presented at *Frontier Detector for Frontier Physics - 15th Pisa Meeting on Advanced Detectors*, La Biodola, 22-28 May 2022.

[25] A. Drago et al., *Ultra-Fast Infrared Detector for Astronomy*, *Nucl.Instrum.Meth.A* Vol.1048, March 2023, 167936. https://doi.org/10.1016/j.nima.2022.167936

[26] A. Drago et al., *Fast Transient Infrared Detection for Time-domain Astronomy*, *JINST* 18 C02012 https://doi.org/10.1088/1748-0221/18/02/C02012. *Journal of Instrumentation*. Part of ISSN: 1748-0221

[27] https://dafne-light.lnf.infn.it/beamlines/sinbad-ir/characteristics/

[28] A. Drago et al., *Performance evaluation of an Ultra-Fast IR Detector for Astronomy Transients*. NANOINNOVATION-2022 Proc., August 1st 2023. Citation : A Drago et al 2023 *J. Phys.: Conf. Ser.* 2579 012013, DOI: 10.1088/1742-6596/2579/1/012013 https://iopscience.iop.org/article/10.1088/1742-6596/2579/1/012013

[29] L. Naponiello et al., L., *Photometry of transients and variable sources at the Osservatorio Polifunzionale del Chianti (OPC)*, in *Proc. SPIE 10704, Observatory Operations: Strategies, Processes, and Systems VII*, 107042C (10 July 2018); https://doi.org/10.1117/12.2313495